\documentclass[conference]{IEEEtran}

\usepackage{cite}
\usepackage{booktabs}
\usepackage{fleqn}
\usepackage{multirow}
\usepackage{amssymb}
\usepackage{amsfonts}
\usepackage{amsmath}
\usepackage{graphicx}

\usepackage{array}
\usepackage{color}
\usepackage{bm}
\usepackage[linesnumbered,algoruled,boxed,lined]{algorithm2e}
\usepackage{siunitx}
\usepackage{graphicx} 
\usepackage{epstopdf}
\usepackage{caption}
\usepackage{graphicx}
\usepackage{float} 
\usepackage{subcaption}
\usepackage{utfsym}
\usepackage{cite}
\usepackage{amsmath,amssymb,amsfonts}
\usepackage[linesnumbered,algoruled,boxed,lined]{algorithm2e}
\usepackage{graphicx}
\usepackage{textcomp}
\usepackage{xcolor}

\makeatletter
\renewcommand\normalsize{%
	\@setfontsize\normalsize\@xpt\@xiipt
	\abovedisplayskip 1pt plus 2pt minus 5pt
	\belowdisplayskip 1pt plus 2pt minus 5pt
	\abovedisplayshortskip 1pt plus 2pt minus 5pt
	\belowdisplayshortskip 1pt plus 2pt minus 5pt
	\let\@listi\@listI}
\makeatother

\ifCLASSINFOpdf
\else
\fi
\begin{document}
	
		%
		\title{UAV Deployment Optimization in UAV-assisted Wireless Communications}
		

		%
		%
		%
		
%
%

%
%
		\author{\IEEEauthorblockN{		
				Xueqi Zhang\IEEEauthorrefmark{2},		
				Aimin Wang\IEEEauthorrefmark{2},
				Geng Sun\IEEEauthorrefmark{2}\IEEEauthorrefmark{3},
				Lingling Liu\IEEEauthorrefmark{2}\IEEEauthorrefmark{4},
                Jing Zhang\IEEEauthorrefmark{5}\IEEEauthorrefmark{4}
			}
			
			\IEEEauthorblockA{\IEEEauthorrefmark{2}College of Computer Science and Technology, Jilin University, Changchun 130012, China}
			\IEEEauthorblockA{\IEEEauthorrefmark{3}Key Laboratory of Symbolic Computation and Knowledge
			Engineering of \\ Ministry of Education, Jilin University, Changchun 130012, China}
   \IEEEauthorblockA{\IEEEauthorrefmark{5}College of Communication Engineering, Jilin University, Changchun 130012, China}
			\IEEEauthorrefmark{4}Corresponding author: Lingling Liu and Jing Zhang}
		%
		%

	\markboth{Journal of \LaTeX\ Class Files,~Vol.~14, No.~8, August~2015}%
	{Shell \MakeLowercase{\textit{et al.}}: Bare Demo of IEEEtran.cls for IEEE Journals}
	%


\maketitle
	
\IEEEpubidadjcol

\begin{abstract}
Due to the fact that the locations of base stations (BSs) cannot be changed after they are installed, it is very difficult to communicate directly with remote user equipment (UE), which will directly affect the lifespan of the system.  Unmanned aerial vehicles (UAVs) offer a hopeful solution as mobile relays for fifth-generation wireless communications due to the flexible and cost-effective deployment. However, with the limited onboard energy of UAV and slow progress in energy storage technology, it is a key challenge to achieve the energy-efficient communication. Therefore, in this article, we study a wireless communication network using a UAV as a high-altitude relay, and formulate a UAV relay deployment optimization problem (URDOP) to minimize the energy consumption of system by optimizing the deployment of UAV, including the locations and number of UAV hover points. Since the formulated URDOP is a mixed-integer programming problem, it presents a significant challenge for conventional gradient-based approaches. To this end, we propose a self-adaptive differential evolution with a variable population size (SaDEVPS) algorithm to solve the formulated URDOP. The performance of proposed SaDEVPS is verified through simulations, and the results show that it can successfully decrease the energy consumption of system when compared to other benchmark algorithms across multiple instances.
\end{abstract}
\begin{IEEEkeywords}
UAV, deployment optimization, differential evolution, variable population size, energy-efficient communication.
\end{IEEEkeywords}
	
%
\IEEEpeerreviewmaketitle
%
%
\section{Introduction}
\label{Introduction}
\par User equipment (UE) has the capacity to collect and analyze data for decision-making \cite{fu2020joint}, and UE can be integrated with other facilities to provide possibilities for numerous emerging applications, such as traffic control and smart homes. In these scenarios, a multitude of devices like mobile phones, monitors, and roadside detectors, are deployed to sense and gather data in specific regions. Moreover, UE is mainly low-power and requires continuous power supply. Therefore, to sustain the uninterrupted operation of the wireless communication network where an UE is located, energy consumption is one of the key constraints.

\par However, it is usually costly to deploy many base stations (BSs) in the areas of interest, and transmitting data from remote UEs to BSs requires multiple hops, which may lead to data loss. Relaying is a key technique in wireless networks, which contributes to extend the coverage area, boost throughput, and reduce fading. Moreover, it comes in two forms that are the fixed (i.e., static relays) and mobile (i.e., dynamic relays) relaying methods \cite{pabst2004relay}. Nowadays, mobile relays are usually low-cost, and have been deployed on vehicles like cars, buses, and trains, which have some limits to the development of these mobile relays because of the limited vehicle movement. However, unmanned aerial vehicles (UAVs) have garnered interest due to their superior mobility, and can establish unobstructed line-of-sight (LoS) links with the height advantage, which further increases the capacity of system. In addition, with their growing mobility and decreasing costs, UAVs are anticipated to have a more crucial role in future wireless communication networks \cite{zeng2019accessing}, therefore we adopt a UAV as a mobile relay to solve the transmission issue between BS and remote UEs. 

\par Despite the numerous benefits of UAVs, UAV-assisted communication in relay network encounters challenges due to their constrained on-board energy. Specifically, UAV requires power not only for communication but also to sustain flight, and battery technology progress is sluggish, which results in a critical limitation on the lifetime of UAVs. Therefore, achieving energy-efficient communication in UAV-assisted relay networks is of paramount importance.

\par To attain the energy-efficient communication, an efficient approach is to optimize the deployment of UAVs. For example, Mozaffari et al. \cite{Mozaffari2017} focus on the optimal arrangement of several UAVs for gathering data from UEs. The authors in \cite{Fan2018} study the arrangement of UAV-assisted relay systems for data transmission. The authors in \cite{fu2020joint} deploy multiple UAVs as relays to enhance the received signal strength of BSs.
Guo et al. \cite{guo2020joint} investigate a UAV-relaying system with the objective of effectively fulfilling user requirements by optimizing transmit power allocation and placement of UAV. The abovementioned studies predetermine the number of UAV hover points, and solely optimize the locations. However, if the predefined number is not optimal, it may lead to a suboptimal deployment \cite{Huang2020}. 

\par Therefore, our work stands out by simultaneously optimizing the number and locations of UAV deployment hover points to decrease the energy consumption of UAV-assisted wireless communication network. 
For traditional gradient-based methods, it is difficult to optimize the above two variables at the same time \cite{huang2020energy}, while evolutionary algorithms (EAs) can solve the above problems with its unique advantages, thereby EAs will be used to solve the problem of this work. EAs operate with a population of individuals, where each individual stands for a potential solution (i.e., the entire deployment). The primary contributions of this research can be outlined as follow.

%
%
%
%
\begin{itemize}
	
    \item We define a UAV relay deployment optimization problem (URDOP) to minimize system energy consumption by optimizing the number and locations of hover points for UAV deployment.
	
    \item To solve the defined URDOP, a self-adaptive differential evolution with a variable population size (SaDEVPS) algorithm is proposed, in which a new coding operator is used to adaptively adjust the population size. A self-adaptive mutation and crossover operator is incorporated into the traditional DE algorithm to maintain the equilibrium between exploration and exploitation capabilities of population. Moreover, the crossover operator on the basis of DE is used to keep the diversity of population. 
	
    \item Simulations are conducted to investigate the effectiveness and performance of the novel proposed SaDEVPS.
	
\end{itemize}	
\par The rest of this article is organized as follows. Section \ref{ SYSTEM MODEL AND PROBLEM FORMULATION} introduces the system model and problem formulation. Section \ref{Proposed Approach} proposes SaDEVPS. Section \ref{Simulation Results} presents the simulation results. The work is concluded in Section \ref{Conclusion}.
	
	%
	%

\section{System Model and Problem Formulation}
\label{ SYSTEM MODEL AND PROBLEM FORMULATION}
	
\subsection{System model}

\par
\begin{figure}
		\centering{\includegraphics[width=3in]{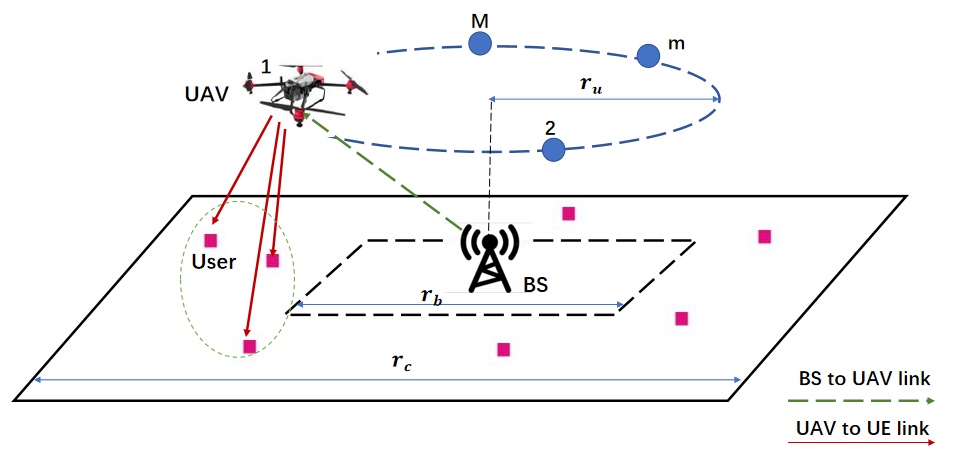}}
		\caption{UAV-enabled mobile relay network topology.}
		\label{System_model}
\end{figure}

 In this work, we study a UAV-assisted relay network as shown in Fig. \ref{System_model}, in which a ground BS is positioned centrally, and multiple remote UEs scattered within an area bounded by a rectangle of inner edge length $rb$ and outer edge length $rc$, and UEs aim to receive data from the BS. 

\par We employ a 3D Cartesian coordinate system, in which the coordinates of a fixed-position BS are represented as $cr_{\text{BS}}$ = (0, 0, 0) and the coordinates of $k$th UE are denoted as $cr_{k}$ = ($x_k$, $y_k$, 0), $k$ = $\{1, 2, ..., K\}$. The UAV flies along a pre-determined trajectory at a fixed altitude $H$ and a steady speed $v$. In addition, UAV can hover at different hover points, and we denote the number of hover points as $M$. The collection of hover points is denoted as $\mathcal{M}$ = $\{1, 2, ..., M\}$. Moreover, at each hover point, the UAV handles data reception and transmission tasks in two phases, that are receiving the signal from BS in the first phase, then decoding and re-encoding signals for transmission to corresponding users in the second phase. The coordinates of the UAV at hover point $m$ are represented as $cr_{\text{UAV}}$ = ($x_{\text{UAV}}^m$, $y_{\text{UAV}}^m$, 0). The distance between the $k$th UE and the UAV is represented as:

\begin{equation}
	d_{k,m} = \sqrt{(x_k-x_{\text{UAV}}^m)^2+(y_k-y_{\text{UAV}}^m)^2+H^2}, 
\end{equation}
\par It is noted that the take-off and landing phases of the UAV are omitted in this study. To minimize the transmission energy, we adopt such a strategy that each UE selects the nearest hover point of the UAV to receive data. In addition, the association between the $k$th UE and UAV at the $m$th hover point{\footnote{Note that there is only one UAV in our work, and the UAV at the $m$th hovering point represents the UAV hovering at the $m$th position, and it is the UAV at hover point $m$.}} is represented by a binary variable $a_{k,m}$, where $a_{k,m}$ = 1 indicates that the data transmission from UAV at the $m$th hover point to the $k$th UE; otherwise, $a_{k,m} = 0$.
\subsection{Energy consumption model of UAV}
\par We consider that the wireless links between UAV and BS, as well as between UEs and UAV are primarily characterized by LoS channels. The channel power gains of BS-UAV and UAV-UE connections, follow free-space path loss model, and can be expressed as\cite{Bastuerk2022a}:
\begin{align}
	& h_{\text{BS},\text{UAV}}^m =  \frac{\beta_0}{(x_{\text{UAV}}^m)^2+(y_{\text{UAV}}^m)^2+H^2}, \\
	& h_{\text{UAV},k}^m = \frac{\beta_0}{(x_k-x_{\text{UAV}}^m)^2+(y_k-y_{\text{UAV}}^m)^2+H^2},
\end{align}

\noindent where $c$ and $f_c$ represent the speed of light and carrier frequency, respectively. The reference channel coefficient at distance $d_0$ = 1 meter is denoted by $\beta_0 = ({4\pi f_c}/{c})^{-2}$. The ratio of channel to noise for two phases at hover point $m$ can also be respectively written as $r_{\text{BS},\text{UAV}}^m = {h_{\text{BS},\text{UAV}}^m}/{\sigma^2}$ and $r_{\text{UAV},K}^m = {h_{\text{UAV},K}^m}/{\sigma^2}$. The noise power over the bandwidth $B$ is denoted as  $\sigma^2 = N_0B$, where $N_0$ represents the noise power spectral density. Therefore, the data rate between transmitter $i$ and receiver $j$ at each hover point $m$ is stated as follows:

\begin{equation}
	R_{i,j}^m = B\log_2(1+P_{i,j}^m r_{i,j}^m),
\end{equation}

\noindent where $i\in\{\text{BS},\text{UAV}\}$ and $j\in \{\text{UAV},k\}$. $P_{i,j}^m$ is the transmit power from transmitter $i$ to receiver $j$ at hover point $m$. We assume that the BS transmits $D_{k}$ amount of data to the $k$th UE, and the time required to transmit this data from $i$ to $j$ is $T_{i,j}^m = D_{k}/R_{i,j}^m$. When UAV receives data from BS, it is mainly the BS that works, therefore the energy consumption of UAV that is used to receive data is very small and can be ignored. Therefore, the energy consumption of UAV is mainly considered when UAV transmits data to the associated UE, and can be structured as:
\begin{equation}
	\begin{aligned}
		E_{\text{UAV}}^t =	\sum_{k=1}^K\sum_{m=1}^M a_{k,m}\cdot P_{\text{UAV},k} \cdot T_{\text{UAV},k}^m.
	\end{aligned}
\end{equation}

\par It moves to the next hover point only after collecting the data sent from BS and sending it to the corresponding UE. Therefore, the hover time at the $m$th hover point consists of two components, that are the time of collecting data and transmitting data. Furthermore, the hover time of UAV at the $m$th hover point can be expressed as:

\begin{equation}
	\begin{aligned}
		T_{h}^m =  \mathop{\max}_{k \in K} \{a_{k,m}\cdot T_{\text{BS},\text{UAV}}^m\}+\mathop{\max}_{k \in K} \{a_{k,m}\cdot T_{\text{UAV},k}^m\}
	\end{aligned}.
	\label{e14}
\end{equation}

\par In this study, not only the communication energy consumption is important, but also the propulsion energy consumption is crucial, which ensures that the UAV stays hover and moves forward. For a rotor UAV traveling at a velocity $v$, the propulsion power of flying in a flat two-dimensional plane is given by \cite{Zeng2019}:
\begin{equation}
	\begin{aligned}
		P(v)= & p_B\left(1+\frac{3 v^2}{v_{\text {tip}}^2}\right)+p_I\left(\sqrt{1+\frac{v^4}{4 v_0^4}}-\frac{v^2}{2 v_0^2}\right)^{1 / 2} \\
		& + \frac{1}{2} d_0 \rho s A v^3,
	\end{aligned}
	\label{ep}
\end{equation}
\noindent where $p_B$ and $p_I$ represent the blade profile power and induced power during hover status, and their values are fixed. $v_0$ represents the average rotor-induced velocity while hovering, and $v_{\text{tip}}$ corresponds to the tip speed of the rotor blades. Parameters $s$ and $d_0$ pertain to rotor solidity and fuselage drag ratio. Additionally, $A$ and $\rho$ are established as the rotor disc area and air density.
\par When setting $v=0$ in Eq. (\ref{ep}), we derive the power consumption for hover status of UAV as $P_h = p_B + p_I$. $P_h$ is finite and contingent on factors such as air density, aircraft weight, and rotor disc area. Since UAV flies around a fixed trajectory with a fixed speed $v$, the flight energy consumption is a constant, and we only consider the hover energy consumption. The total hover energy consumption of UAV is given by:
\begin{equation}
	\begin{aligned}
		E_{\text{UAV}}^h=\sum_{m=1}^M P_h T_h^m.
	\end{aligned}
\end{equation}
\vspace{-0.8cm}
\subsection{Problem formulation}
\par We consider such a relay scenario as shown in Fig. \ref{System_model}, where the BS first transmits data to UAV, and UAV transmits the collected data to the corresponding UE by hovering at different hovering points.

\par According to the above description, the energy consumption of UAV is mainly genrated in the second stage, including the energy of hovering and transmitting data to UE. When transmitting data to UE, the distance between UE and UAV influences the transmission rate, which further determines the energy consumption of UAV. Furthermore, the distance is mainly affected by the number of hover points and position of the UAV. The fewer hover point means that the distance between UE and UAV is farther, resulting in a lower transmission rate, and the UAV requires more energy to complete transmission. On the contrary, although the distance between UE and UAV is shorter if the number of hover points is more, there may be no UE that needs to receive data at the point where the UAV is hovering. Therefore, adaptively adjusting the number and locations of hover points is a key factor that having an impact on the UAV energy consumption. We will comprehensively consider these factors and formulate a URDOP as follows:
\begin{subequations}
	\label{problem}
	\begin{align}
		&{\mathop{\min}_{{\{x_{\text{UAV}}^m,y_{\text{UAV}}^m},M\}} E_{\text{UAV}}^h + \phi E_{\text{UAV}}^t} \\	 
		&\qquad\sum_{m=1}^M a_{k,m} = 1, \quad\forall k \in K\\
		&\qquad\sum_{k=1}^K a_{k,m} \le N,\quad \forall m \in M\\
		&\qquad	\sum_{k=1}^K\sum_{m=1}^M a_{k,m} = K\\
		&\qquad M_{min} \le M \le M_{max}
	\end{align}
\end{subequations}
\noindent where $\phi \geq 0$ represents the weight between UAV hover energy consumption and data transmission energy consumption. Eq. (\ref{problem}b) enforces that each UE can communicate with only one UAV hover point. To account for system bandwidth constraints, UAV can simultaneously serve up to $N$ UEs at each hover point, as indicated in Eq. (\ref{problem}c). Additionally, Eq. (\ref{problem}d) ensures that all UEs receive service. $M_{min}$ and $M_{max}$ represent the lower and upper bounds on the number of hovering points $M$.

\par Note that in this investigation, in order to facilitate energy-efficient communication, we mainly emphasize optimizing the deployment strategy of UAV, including the number and locations of hover points for UAV flying along a fixed trajectory~\cite{li2020throughput}.

\section{Proposed Approach}
\label{Proposed Approach}
\subsection{Motivation}
\par
\begin{figure}[htb]
	\centering{\includegraphics[width=2.5in]{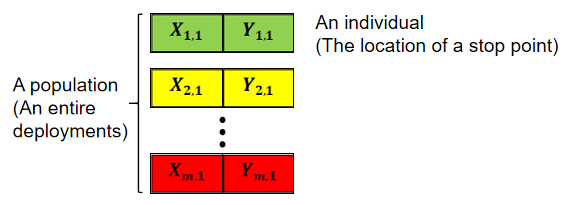}}
	\caption{The encoding mechanism.}
	\label{c}
\end{figure}

 In order to solve the formulated URDOP, the number and locations of hover points are optimized. The number of hover points is dynamically changed, which indicates that the gradient vector is variable during the optimization process, making it intractable to be solved by using traditional methods. EAs have potential to solve the problem due to the nature of gradient-free, in which DE is one of the most widely used EAs working with a group of individuals and each individual represents a candidate solution. To solve the dynamic solution length problem, Huang et al. \cite{Huang2020} introduce a new encoding mechanism, as shown in Fig. \ref{c}, in which each individual encodes the coordinates of hover points. Moreover, the population size represents the total hover points and is optimized during evolutionary process. Additionally, the positional dimension of each hover point is simplified to 2 (i.e., $x$-axis and $y$-axis), and the flight height $H$ remains constant, which reduces the search space dimension. However, the evolutionary process in \cite{Huang2020} lacks optimization, employing only a mutation operator and a fixed crossover operator. Therefore, it is important to improve DE to make it suitable for solving the above problems.

\vspace{-0.2cm}
\subsection{Standard DE}
\label{DIFFERENTIAL EVOLUTION (DE)}	

\par DE is a robust stochastic search method for global optimization \cite{Storn1997}. It begins with the random generation of an initial population, where each individual is termed as a target vector:
\begin{equation}
	\begin{aligned}
		{x_i=(x_{i,1}, x_{i,2}, ..., x_{i,D}), i=\{1, 2, ..., {NP}\}},
	\end{aligned}
\end{equation}

\noindent where $D$ stands for the count of variables, and $NP$ represents the population size. Following the initialization, a mutation operator is applied into the target vector $x_i$, which results in the creation of a mutant vector $v_i$. Subsequently, a trial vector $u_i$ is derived through the implementation of a crossover operator on $v_i$ and $x_i$. Lastly, a selection operator determines the superior candidate between $u_i$ and $x_i$ to advance to the next generation. Following, we will present the mutation, crossover, and selection operators.
\begin{itemize}
	\item [1)]
	Mutation: Frequently employed mutation operators are: \\
	$\bullet$ DE/rand/1
	\begin{equation}
		\begin{aligned}
			v_i = x_{r0} + F \cdot (x_{r1} - x_{r2}),
		\end{aligned}
		\label{e19}
	\end{equation}
	$\bullet$ DE/rand/2
	\begin{equation}
		\begin{aligned}
			v_i = x_{r0} + F \cdot (x_{r1} - x_{r2}) + F \cdot (x_{r3} - x_{r4}),
		\end{aligned}
		\label{e20}
	\end{equation}
 
\noindent where the indices $r0, r1, r2, r3$, and $r4$ are unique integers freely selected from $[1, {NP}]$, which are different from $i$. $F$ is the scaling factor.
\item [2)]
Crossover: The binomial crossover is given as follows

	\begin{equation}
		\label{crossover}
		\begin{aligned}
				u_{j,i}=
			\begin{cases}	
				v_{j,i}, & \text{if} ~rand(0,1) \le CR ~\text{or}~ j=j_{rand},\\
				x_{j,i}, & \text{otherwise},
			\end{cases}
		\end{aligned}
	\end{equation}

\noindent where $j_{rand}$ is a freely selected integer from the range of [1, $D$]. $rand$ represents a random number uniformly allocated in (0,1), and the crossover probability is denoted by $CR$ $\in$ (0,1).
\item [3)]Selection: In a minimization issue, the selection operator is executed as follows:

\begin{equation}
	\label{select}
	\begin{aligned}
		x_{i}=
		\begin{cases}	
			u_{i}, & \text{if $f(u_{i}) \le f(x_{i})$},\\
			x_{i}, & \text{otherwise}.
		\end{cases}
	\end{aligned}
\end{equation}		

\end{itemize}


\subsection{SaDEVPS}
\addtolength{\topmargin}{0.01in}
\par As shown in Algorithm \ref{General framework of SaDEVPS}, the first step is to create a population $P$ randomly within the search space. SaDEVPS then uses the self-adaptive mutation and crossover operator to generate offspring population $Q$, which is subsequently used to update $P$ to the next generation. The process is iterated until the stop condition is satisfied, which is defined as the maximum number of fitness evaluations ${MaxFE}$. In addition, the new coding operator and initialization of population $P$, self-adaptive mutation and crossover operator, crossover operator of DE, and generation of offspring population are described in detail below.
\begin{algorithm}
	\caption{General framework of SaDEVPS}
	\label{General framework of SaDEVPS}
	$FE$ = 0; \\
	Initialize $P$; \\
	\While{$FE \le MaxFE$}{$Q$=$\emptyset$; \\ 
		\For{$i=1$ to $\left|P\right|$}{Apply 'DE/rand/1', 'DE/rand/2', and the binomial crossover on $x_i$ in $P$ to produce $u_i$; $Q=Q \cup u_i$; \\
		
		}
		Updating $P$ according to Algorithm \ref{Updating of population P};}
\end{algorithm}

\par \emph{\textbf{1)} {New coding operator and initialization of population $P$:}} In SaDEVPS, the population size of $P$ aligns with the number of hover points, and it dynamically optimize the quantity. The key concept of this paper is to modify at most one individual in $P$ during each update, which can dynamically adjust the population size and ensure a stable search space. 

\par We begin by randomly generating the locations of $M_{max}$ hover points to create the initial population $P$. With each initialization of $P$, the $FE$ ($FE$  represents the number of fitness evaluations) increases by one. Afterwards, we check whether $P$ satisfies all constraints given in Eq. ({\ref{problem}}) to determine whether it is feasible. If $P$ is feasible, we have successfully generated the initial UAV deployment. Otherwise, we repeat the above process until either $P$ is feasible, or \emph{FE} $\ge$ \emph{MaxFE}.

%
	
\par \emph{\textbf{2)} {Self-adaptive mutation and crossover operator:}} To produce a new population $Q$, the self-adaptive learning strategy probabilistically select one of the two learning approaches (i.e., `DE/rand/1' in Eq. (\ref{e19}) and `DE/rand/2' in Eq. (\ref{e20})), and applies it into the current population $P$. The selection probability of using the `DE/rand/1' strategy for each individual is denoted as $p_1$, while the probability for another strategy is $p_2 = 1-p_1$. Initially, the probability of each individual is equal (i.e., $p_1 = p_2$). If $p$ is less than or equal to $p_1$, the `DE/rand/1' strategy is employed; otherwise, the `DE/rand/2' strategy is used. After assessing the newly generated trial vectors, the counts of successfully entering into the next generation for `DE/rand/1' and `DE/rand/2' strategies are respectively denoted as $s_1$ and $s_2$. The counts of trial vectors discarded for each strategy are labeled as $f_1$ and $f_2$, respectively. The counts are amassed over a defined number of generations, referred to as the learning period ($LP$). Subsequently, $p_1$ and $p_2$ are recalculated as follows:
\begin{equation}
\begin{split}
    &p_1=\frac{s_1\cdot(s_2+f_2)}{s_2\cdot(s_1+f_1)+s_1\cdot(s_2+f_2)},\\
    & p_2=1-p_1,
    \label{p1}
\end{split}
\end{equation}
\noindent where the probability of devoting these two strategies is recalculated after $LP$. $s_1, s_2, f_1$, and $f_2$ are reset at each update to avoid possible side effects that are accumulated during the previous learning phase. The equilibrium between the exploitation and exploration capabilities of the population has been kept after using this operation.
\par \emph{\textbf{3)} {Crossover operator of DE:}} In DE, the control parameters $CR$ and $F$ are critical, which follow normal distribution. We record the values of $CR$ associated with the test vectors for each generation, that is successfully advanced to next generation as $CRs$. After a defined number of generations (i.e., the abovementioned $LP$), we recalculate the mean $CR_a$ based on the values stored in $CRs$. It is worth noting that after recalculating the mean, the data in $CRs$ is removed to prevent potentially inappropriate cumulative effects over time. We incorporate the aforementioned learning strategy and parameter self-adaptive scheme into the DE evolutionary process.

%

\begin{algorithm}
	\caption{Updating of population $P$}
	\label{Updating of population P}
	\For{$i=1$ to $|Q|$}{
		$P_1 \gets$ Incorporate the $i$th individual from $Q$ into $P$;\\
		$P_2 \gets$ Employ the $i$th individual from $Q$ to substitute a freely chosen individual in $P$;\\
		$P_3 \gets$ Remove a freely chosen individual from $P$;\\
		\eIf{at least one of the populations $P_1$, $P_2$, and $P_3$ is feasible}{
			\If{the performance is improved against $P$}{
				The feasible population with the most significant performance is employed to substitute $P$; \\
				Update the values of $CRs$, $s_1$, $f_1$, $s_2$, and $f_2$ ;
			}
		}{
			\If{the fitness value of feasible population $P_3$ is equal to that of $P$}{$P$ is replaced by $P_3$;}
		}
		$FE = FE +3$; \\
	}
\end{algorithm}

\par \emph{\textbf{4)} {Generation of offspring population:}} Following the generation of offspring population $Q$, two new populations are generated by using the individuals in $Q$ according to the following procedures.

\begin{itemize}
	\item The first individual in $Q$ is added into $P$ to produce a new population $P_1$.
	\item An arbitrarily selected individual from $P$ is replaced by the first individual in $Q$ to form a new population $P_2$.
\end{itemize}

\par The third new population $P_3$ is generated when randomly selecting an individual from $P$ and removing it.

\addtolength{\topmargin}{0.03in}
\par We then assess the feasibilities of $P_1$, $P_2$ and $P_3$. If at least one of these populations is feasible, we evaluate its fitness value. Subsequently, the population with the highest performance enhancement replaces $P$. If there is no improvement, and the fitness value of $P_3$ matches the fitness value of $P$, then $P$ is replaced by $P_3$, which indicates that the removed individual has no impact on performance. The remaining individuals in $Q$ perform the above operations one by one, and the specific process is shown in Algorithm \ref{Updating of population P}.
\vspace{-0.1cm}
\section{Simulation Results}
\label{Simulation Results}
\par In this section, we employ Matlab to conduct simulations and assess the performance of the proposed SaDEVPS in solving the formulated URDOP. Furthermore, we compare the performance of SaDEVPS with the following four algorithms.
\begin{itemize}
	
        \item MTDE integrates a self-adaptive motion step from the novel multi-trial vector (MTV) approach, amalgamating various search strategies via a trial vector producer (TVP). The number of hover points in MTDE is predetermined and remains constant throughout the evolution\cite{NadimiShahraki2020}.
	
	\item fGA operates in phenotype space, swaps two circular regions of equal size from two square regions, and subsequently creates the two new square regions \cite{Zhang2017}.
	
	\item DEVIPS applies a novel coding approach, that treats the entire population as an entire deployment \cite{Huang2020}.
	
	\item DEEM uses a coding mechanism, that is similar to DEVIPS. Similar to MTDE, it necessitates a predetermined quantity of hover points \cite{Wang2018}.
	
\end{itemize}	

\begin{table}[htb]   
	\begin{center}   
		\caption{Parameter settings}  
		\label{table:1} 
		\begin{tabular}{|c|c|}   
			\hline   Algorithm & Parameter Settings \\   
			\hline   DEEM & $F$ = 0.9, $CR$ = 0.9\\ 
			\hline   MTDE &$NP$ = 100, Winlter = 20, H = 5\\  
			\hline   fGA & $NP$ = 100, $p_c$ = 0.9, $p_m$ = 0.1 \\      
			\hline   DEVIPS & $F$ = 0.6, $CR$ = 0.5 \\  
			\hline \multirow{2}{*}{SaDEVPS} & $p_1=p_2$=0.5, $CR_a$=0.5, $LP$=50, \\ &$F\sim N(0.5,0.3)$, $CR\sim N(CR_m,0.1)$ \\     
			\hline   
		\end{tabular}   
	\end{center}   
\end{table}
	\begin{figure*}[htbp]
	\centering
	\begin{subfigure}{0.18\linewidth}
		\centering
		\includegraphics[width=1.1\linewidth]{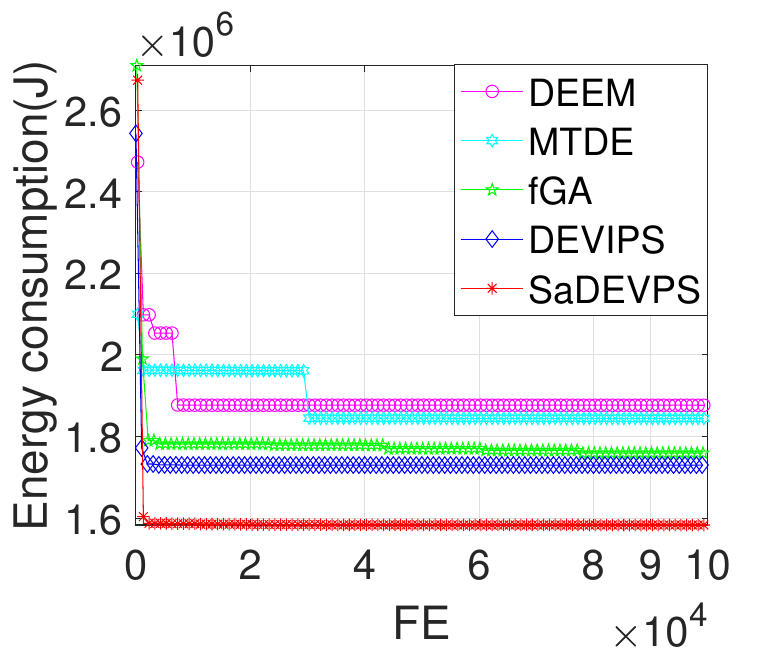}
		\caption{$K$=100}
	\end{subfigure}
	\centering
	\begin{subfigure}{0.18\linewidth}
		\centering
		\includegraphics[width=1.1\linewidth]{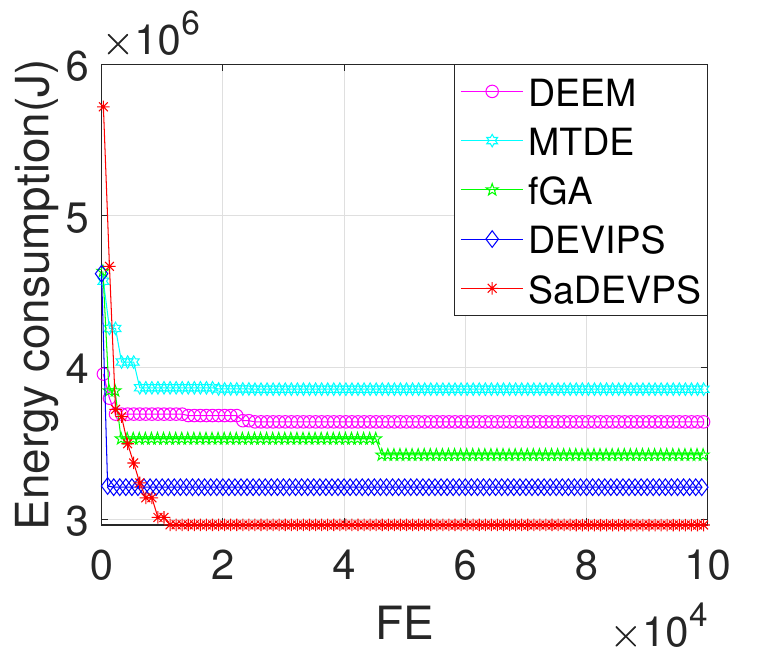}
		\caption{$K$=200}
	\end{subfigure}
	\centering
	\begin{subfigure}{0.18\linewidth}
		\centering
		\includegraphics[width=1.1\linewidth]{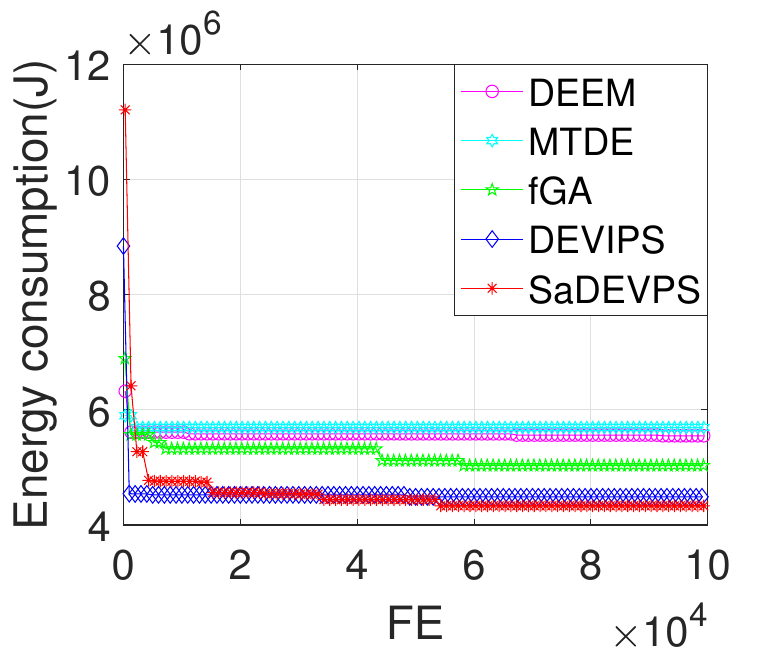}
		\caption{\centering{$K$=300}}
	\end{subfigure}
	\centering
	\begin{subfigure}{0.18\linewidth}
		\centering
		\includegraphics[width=1.1\linewidth]{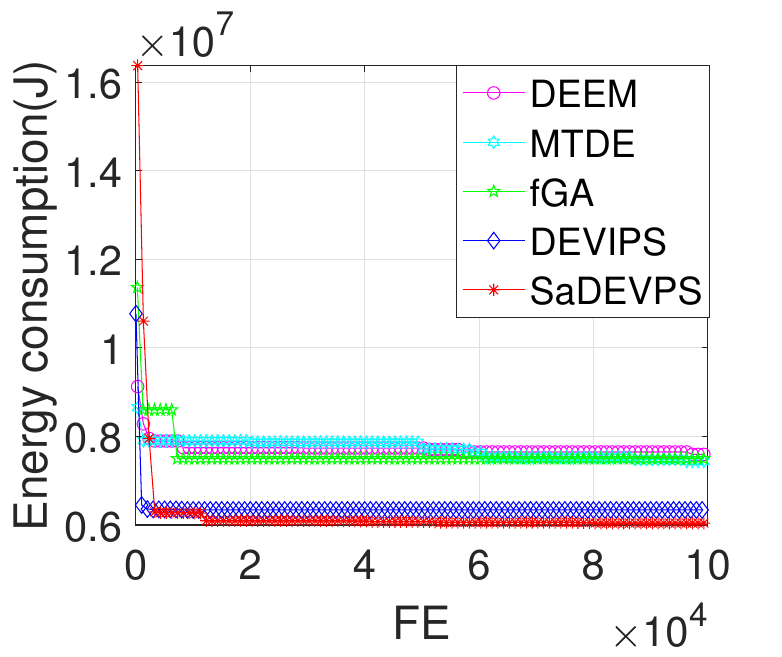}
		\caption{\centering{$K$=400}}
	\end{subfigure}
	\centering
	\begin{subfigure}{0.18\linewidth}
		\centering
		\includegraphics[width=1.1\linewidth]{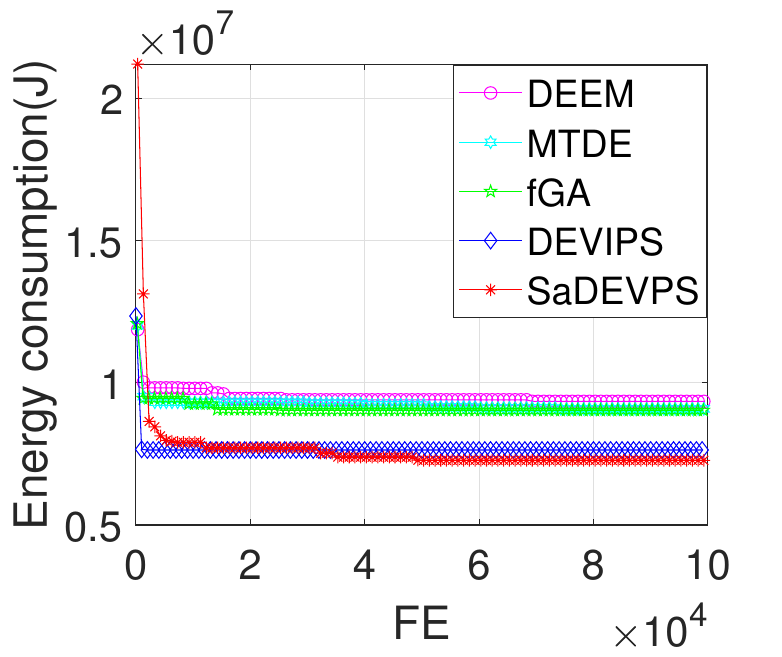}
		\caption{$K$=500}
	\end{subfigure}
	\caption{Evolution of the mean energy consumption obtained by SaDEVPS and other four approaches.}
	\label{compare}
\end{figure*}
\vspace{-0.6cm}	
\subsection{Parameter settings}
\par The parameter settings for the four comparative algorithms and SaDEVPS are listed in Table \ref{table:1}.

\begin{table*}[t]
	\centering
	\caption{The mean of energy consumption obtained by SaDEVPS and other four approaches.}
	\label{result}
	\resizebox{\textwidth}{!}{
		\begin{tabular}{|c|c|c|c|c|c|c|c|}
			\hline
			$K$& \multicolumn{2}{c|}{DEEM (J)}&\multicolumn{2}{c|}{MTDE (J)} & \multirow{2}{*}{fGA (J)}& \multirow{2}{*}{DEVIPS (J)} & \multirow{2}{*}{\textbf{SaDEVPS} (J)} \\   
			\cline{2-5}
			UE & $(M_{min}+M_{max})/2$ & $M_{max}$&  $(M_{min}+M_{max})/2$ & $M_{max}$ & & & \\
			\hline 
			100  & 1.8765E+6(15.64$\%$) &1.8968E+6(16.54$\%$) &  1.8304E+6(13.51$\%$) &1.8907E+6(16.27$\%$) & 1.7591E+6(10.01$\%)$ & 1.7300E+6(8.50$\%$)  & 1.5830E+6\\ 	
			\hline
			200  & 3.6418E+6(18.65$\%$) &3.8638E+6(23.33$\%$) &  3.8572E+6(23.20$\%$) & 4.0838E+6(27.46$\%$) & 3.4238E+6($13.48\%$) & 3.2114E+6($7.76\%$) & 2.9623E+6 \\   	
			\hline
			300  &5.5479E+6(21.91$\%$) &5.7240E+6(24.32$\%$) &  5.6907E+6(23.87$\%$) &5.8572E+6(26.04$\%$) &  5.0329E+6($13.93\%$) & 4.4888E+6($3.50\%$) & 4.3319E+6  \\  	
			\hline
			400  & 7.5951E+6(19.77$\%$) & 7.6552E+6(20.40$\%$) & 7.4373E+6(18.07$\%$) & 7.6859E+6(20.72$\%$) &  7.4986E+6($18.74\%$) & 6.3398E+6($3.90\%$) & 6.0931E+6  \\      
			\hline
			500 & 9.3469E+6(22.19$\%$) &9.4604E+6(23.12$\%$) & 9.0831E+6(19.93$\%$) & 9.5324E+7(23.70$\%$) & 9.0344E+6($19.50\%$) & 7.6371E+6($4.78\%$) & 7.2723E+6  \\      
			\hline	
		\end{tabular}
	}
\end{table*}

\par In our simulations, each algorithm will terminate after 100,000 fitness evaluation (i.e., $MaxFE$ = 100,000), and each algorithm undergoes 30 independent runs for each instance, furthermore these results are the mean of these iterations. It is assumed that UAV flies around a circular trajectory of radius $r_u$ centered at (0, 0, 0), where $r_u$ is set to 450 m, UEs are located within a rectangle area with an inner edge length $rc$, and an outer edge length $rb$, in which $rc$ is set to 1500 m and $rb$ is set to $rc/2$. The flight altitude of UAV is 100 m. In addition, $D_i (i \in K)$ is distributed at random within [1, $10^3$] MB, and $N$ is set to 20. The maximum and minimum value for the number of hover points are respectively set to $M_{max}=K$ and $M_{min}=K/N$. $P_{\text{BS}, \text{UAV}}= 1$ W and $P_{\text{UAV},k} (k \in K)=0.1$ W. The carrier frequency $f_c$ and communication bandwidth $B$ are respectively set to 2 GHz and 1 MHz. The spectral density value of noise power is set to $N_0$ = -174 dBm/Hz, the parameters used in calculating $P_h$ are explained in \cite{Zeng2019}, and $\phi=1000$. In addition, the performance of SaDEVPS is evaluated across 5 instances with the varying number of UES, i.e., $K$ = $\{100, 200, 300, 400, 500\}$.
\subsection{Optimization results}
\par The performance of SaDEVPS is compared with the performance of DEEM, MTDE, fGA and DEVIPS. DEEM and MTDE require a predetermined number of hover points, while fGA, DEVIPS and SaDEVPS self-adaptively determine the number of hover points during optimization. For DEEM and MTDE, we test two constant number of hover points, that are $M_{max}$ and $(M_{min} + M_{max})/2$ (i.e., $K$, and $\lfloor \frac{K(N+1)}{2N} \rfloor$). 

\par Table \ref{result} displays the mean of energy consumption obtained from the five algorithms over 30 runs. Fig. \ref{compare} gives the evolution trend of the mean energy consumption obtained from the five algorithms. As can be seen from the Table \ref{result} and Fig. \ref{compare} that, SaDEVPS demonstrates superior performance in energy consumption compared to the other four algorithms across all instances. Specifically, as the number of UE increases, SaDEVPS demonstrates a growing advantage over fGA. Additionally, when compared to DEVIPS, DEEM, and MTDE, SaDEVPS shows a consistent performance improvement with the increasing number of UE.

\par This observation can be elucidated as follows, fGA operates in a dynamically changing dimensional search space as the number of UE increases. SaDEVPS maintains a consistently low and fixed dimension in solution space, which leads to more favorable outcomes. Although DEVIPS also maintains a fixed dimension for search space, SaDEVPS surpasses it by adopting the self-adaptive mutation and crossover operator during population generation, which enhances the efficiency of traditional DE algorithm. To sum up, the excellent performance of SaDEVPS demonstrates the effectiveness of simultaneously optimizing the number and locations of hover points to achieve energy-efficient communication.
\section{Conclusion}
\label{Conclusion}
\par In this study, we investigate a wireless communication system employing a UAV as a mobile relay. First, a URDOP is formulated to minimize the energy consumption of system by concurrently optimizing the number and locations of hover points of UAV. Then, for the purpose of solving the formulated URDOP, a SaDEVPS method is proposed, in which a new coding operator, a self-adaptive mutation and crossover operator, and crossover operator on the basis of DE algorithm are introduced. Simulation results show that the proposed SaDEVPS has the best performance comparing to other optimization methods.
\vspace{-0.3cm}
	\ifCLASSOPTIONcaptionsoff
	\newpage
	\fi

	
	
	%

	\bibliographystyle{ieeetr}
	\bibliography{document}

\end{document}